\documentclass[prl,twocolumn,showpacs,amsmath,letter]{revtex4}
\usepackage{graphicx, verbatim}
\newcommand{\Journal}[4]{#1 \textbf{#2}, #3 (#4)}
\usepackage{CJK}

\begin{document}
\begin{CJK*}{GB}{gbsn}
\title{Spectral characteristics of the microwave emission by the spin Hall nano-oscillator}
\author{R. H. Liu(ÁõÈÙ»ª)}
\author{W. L. Lim}
\author{S. Urazhdin}
\affiliation{Department of Physics, Emory University, Atlanta, GA 30322, USA.}

\begin{abstract}
We utilized microwave spectroscopy to study the magnetization oscillations locally induced in a Permalloy film by a pure spin current, which is generated due to the spin Hall effect in an adjacent Pt layer. The oscillation frequency is lower than the ferromagnetic resonance of Permalloy, indicating that the oscillation forms a self-localized non-propagating spin-wave soliton. At cryogenic temperatures, the spectral characteristics are remarkably similar to the traditional spin-torque nano-oscillators driven by spin-polarized currents. However, the linewidth of the oscillation increases exponentially with temperature and an additional peak appears in the spectrum below the ferromagnetic resonance, suggesting that the spectral characteristics are determined by interplay between two self-localized dynamical states.
\end{abstract}


\maketitle
\end{CJK*}

Spin-polarized electric current can be utilized to modify the magnetic configuration of ferromagnets and is essential for the operation of active spin-based electronic (spintronic) devices~\cite{prinz}. These devices typically utilize two ferromagnetic layers~\cite{katine}. The electric current is spin-polarized in one layer and injected into another (free) layer, changing the magnetic configuration or causing microwave-frequency precession of the latter due to the spin transfer torque (ST) effect~\cite{slon1,berger,tsoi1,kiselev}. The ST exerted by each transmitted electron is generally limited by its total spin-angular momentum $\hbar /2$. As a consequence, device operation requires a relatively large electric current $I$ that scales with the magnetic moment of the free ferromagnet~\cite{slon1}. On the other hand, thermal stability of nanoscale magnetic devices generally improves with increasing magnetic moment~\cite{silva}. Because of these opposite trends, it has been challenging to minimize the current required for the operation of spin torque devices while maintaining their stability.

In a recently developed class of spintronic devices, the spin Hall effect (SHE)~\cite{Diakonov,Hirsch} produces a pure spin current controlling the magnetization of ferromagnets~\cite{Ando_SHE,Liu_SHE,Demidov_SHE,wang,liuprl2}. The efficiency of SHE devices (the angular momentum transferred to the ferromagnet by each electron transmitted through the device) is not limited by the magnitude of the angular momentum of electron, since a single electron  can experience multiple scattering both in the ferromagnet and in the SHE material, transferring a sizable fraction of its total angular momentum in each scattering event. Moreover, since electric current does not need to flow through the active magnetic layer, SHE devices can utilize not only conducting but also dielectric magnetic materials.

While the use of SHE opens possibilities for new device geometries, it also requires new approaches to signal generation and a better understanding of the effects of geometry on the dynamical characteristics of magnetic systems. For instance, it has been demonstrated that magnetization oscillations cannot be induced by a uniform spin current applied to a micron-sized ferromagnetic disk because of the nonlinear damping~\cite{Demidov_SHE}.

Auto-oscillation induced by SHE has been recently observed by micro-focus Brillouin light spectroscopy (BLS) in an in-plane point-contact geometry. By locally injecting a spin current into an extended magnetic film~\cite{Demidov_SHO}, radiative damping of the propagating spin-wave modes was enhanced, suppressing nonlinear processes and enabling auto-oscillation.
Spatial mapping and spectroscopic analysis  showed that the oscillation formed a non-propagating self-localized spin-wave soliton (a spin-wave "bullet") at a frequency below the linear spin-wave spectrum of the magnetic material, confirming the earlier theoretical prediction~\cite{slavinprl}. Ref.~\cite{Demidov_SHO} demonstrated a route for achieving oscillations due to SHE, but did not show how these oscillations can be converted into microwave signals. The spectroscopic resolution of BLS was also insufficient to determine the spectral properties of the oscillation.

Here, we demonstrate coherent microwave generation due to the SHE in a device that utilizes local current injection to suppress nonlinear damping, and anisotropic magnetoresistance (AMR) of the magnetic layer to convert the oscillations into a microwave signal. Our spectroscopic measurements directly demonstrate the coherent single-mode nature of the excitation and confirm its relation to the spin-wave spectrum observed in Ref.~\cite{Demidov_SHO}.
Furthermore, we show that while at low temperatures the linewidth is comparable to other magnetic nano-oscillators, at higher temperatures it significantly increases and an additional peak appears in the spectrum, suggesting a greater complexity of the dynamical states induced by spin current than previously theoretically predicted~\cite{slavinprl}.

\begin{figure}[h]
\centering
\includegraphics[width=0.485\textwidth]{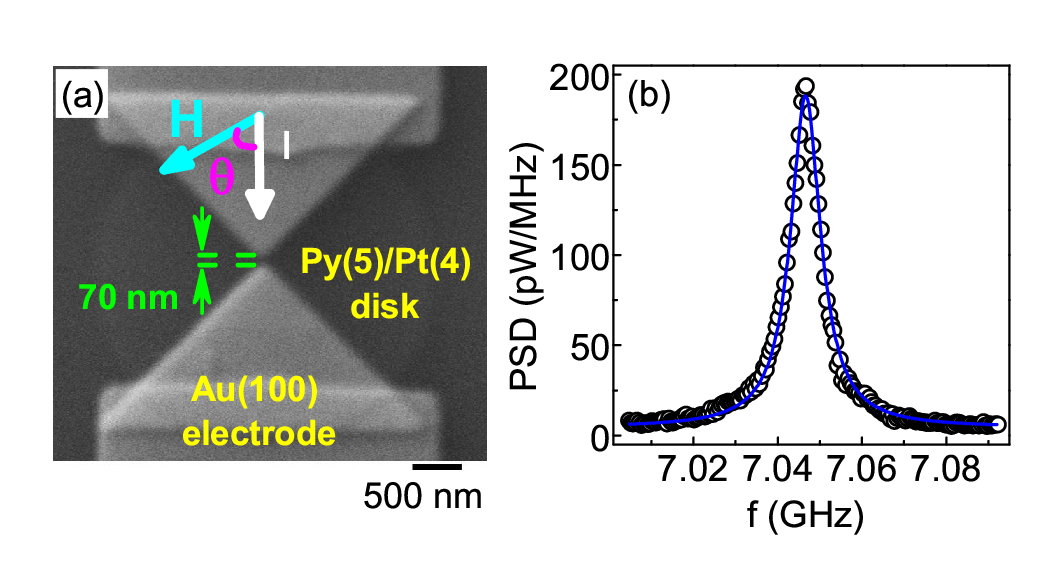}
\caption{(Color online). (a) Scanning electron micrograph of the test device. (b) Symbols: power spectral density (PSD) of the microwave signal emitted by the device at $H = 700$~Oe, $I = 20$~mA, $T=6$~K,  at an angle $\theta=60^o$ between the direction of the field and the current flow. The curve is the result of fitting by the Lorentzian function.}\label{fig1}
\end{figure}

Figure~\ref{fig1}(a) shows a scanning electron microscopy (SEM) image of our test device and the experimental layout. The device is comprised of a $4\ \mu$m Py(5)/Pt(4) bilayer disk and two pointed Au(100) electrodes on top.
Here, Py is Permalloy (Ni$_{80}$Fe$_{20}$), and thicknesses are in nanometers.
The device was fabricated by a combination of sputtering and e-beam lithography.
The separation between the endpoints of the Au electrodes was $70$~nm. By applying a voltage between the Au electrodes, an in-plane electrical current localized mostly in the gap between electrodes was induced in the Py/Pt bilayer. This current induced a pure spin current flowing towards the Py layer due to the SHE in Pt~\cite{Diakonov,Hirsch}, resulting in the oscillation of the Py magnetization ${\mathbf M}$.

To enable microwave generation due to this oscillation, we utilized the dependence of the device resistance $R$ on the angle $\theta$ between ${\mathbf M}$ and the direction of current $I$. Magnetometry measurements showed that the coercivity of the Py film did not exceed $5$~Oe, significantly smaller than the magnitude of the applied in-plane magnetic field $H\ge 180$~Oe in all of our measurements. Therefore, the equilibrium orientiation of ${\mathbf M}$ was parallel to the field ${\mathbf H}$. The sinusoidal dependence of $R$ on the orientation of ${\mathbf H}$ with a period of $180^\circ$ was consistent with the anisotropic magnetoresistance (AMR) of Py~\cite{O'Handley}. The relative magnetoresistance was $\Delta R/R = [R(0^\circ)-R(90^\circ)]/R(90^\circ)=0.125$\% at RT, which was smaller than in standalone Py films due to the shunting by Pt.

Previous studies showed that the current-induced excitation is most efficient when the magnetization, the direction of the current flow, and the normal to the film surface directed from Py to Pt form a right-hand set of orthogonal vectors~\cite{Demidov_SHE, liuprl2}, consistent with the expected symmetry of SHE~\cite{Diakonov,Hirsch}. On the other hand, the AMR has a minimum when ${\mathbf M}$ is orthogonal to $I$, and therefore a microwave signal at the oscillation frequency cannot be produced in this configuration. This limitation of AMR can be avoided by choosing intermediate orientations of ${\mathbf H}$ such that the oscillation of ${\mathbf M}$ is induced by the spin torque, and simulataneously a microwave signal is generated at the frequency of the oscillation. We observed oscillations in the range of $\theta$ between $30^\circ$ and $85^\circ$, as illustrated in Fig.~\ref{fig1}(b) for $\theta=60^\circ$. The spectrum can be well fitted by the Lorentzian function with a full width at half maximum (FWHM) of $5$~MHz at $T=6$~K [solid curve in Fig.~\ref{fig1}(b)]. The actual temperature of the active device area was $T_a\approx 50$~K because of the Joule heating, as discussed below.

\begin{figure}[htbp]
\centering
\includegraphics[width=0.485\textwidth]{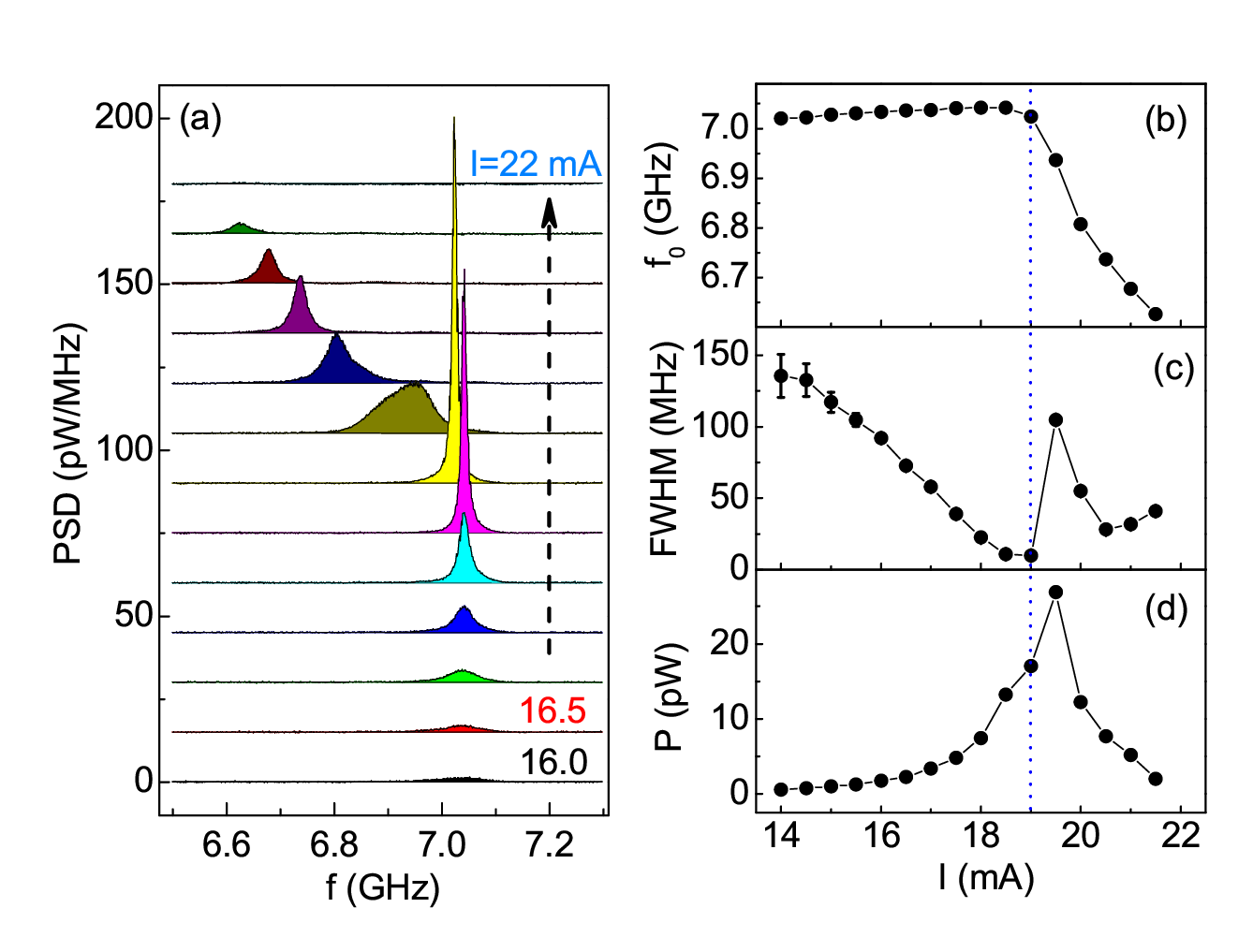}
\caption{(Color online). Dependence of the microwave generation characteristics on current, at $H=700$~Oe, $\theta=60^\circ$, $T=120$~K. (a) Generation spectra at $16$~mA$<I<22$~mA varied in $0.5$~mA increments. (b)-(d) Dependence of the central generation frequency (b), FWHM (c) and integral intensity (d) on current. The central frequency and the linewidth were determined by fitting the power spectra with the Lorentzian function. Dotted vertical line marks the current $I_p$ defined in text.}\label{fig2}
\end{figure}

Figure~\ref{fig2} shows the dependence of the generation characteristics on current $I$ at $T=120$~K. A small oscillation peak appeared at $I>14$~mA. The intensity of the peak quickly increased to a maximum amplitude of $110$~pW/MHz at $I=19$~mA and the frequency slightly increased, while the linewidth decreased to a minimum value of $10$~MHz at $I=19$~mA. At $I>19$~mA, the peak broadened and decreased in amplitude while shifting to lower frequencies. This evolution of the oscillation characteristics is remarkably similar to the traditional spin-valve nano-oscillators both in the nanopillar~\cite{mistral} and point contact geometries~\cite{hysteresis,parametric}. It is consistent with the theory of nonlinear oscillators, which predicts that the thermal linewidth decreases with increasing oscillation power and increases with increasing nonlinearity~\cite{kim1, nonlin-oscil}. In particular, a significant broadening of the oscillation peak at $I>19$~mA is correlated with a strong oscillation redshift at these currents.

\begin{figure}[htbp]
\centering
\includegraphics[width=0.33\textwidth]{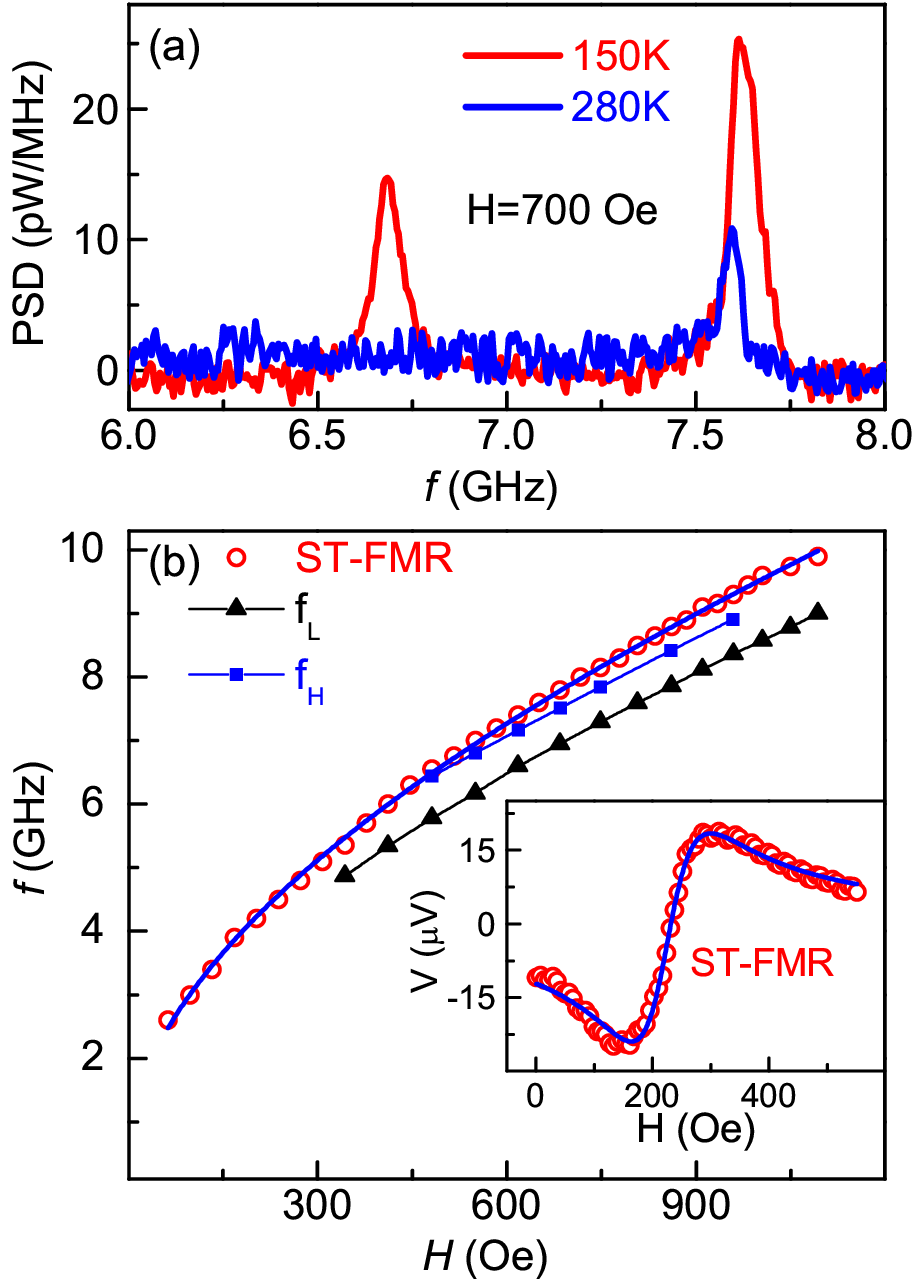}
\caption{(Color online). Relation of the oscillation peaks to the ferromagnetic resonance (FMR) of the Py film. (a) Microwave generation spectra at \emph{T} = 150 K, \emph{I} = 25 mA and \emph{T} = 280 K, \emph{I} = 24 mA. (b) The ST-FMR frequency (circles), the maximum frequency of the low-frequency peak (triangles) and the high-frequency peak (squares) {\it vs} $H$. The solid curve is the result of fitting the FMR data with the Kittel formula $f=\gamma\sqrt{H(H+4\pi M)}$, where $M = 827$~G is the best-fit value of the Py magnetization, and $\gamma = 2.8$~MHz/Oe is the gyromagnetic ratio. Inset: the ST-FMR voltage {\it vs} field $H$ obtained with a mw current of $2.5$~mA rms at frequency $f_{ext} = 4.5$~GHz. The curve is the best fit with a sum of a symmetric and an antisymmetric Lorentzian. $f_{FMR}$ and $f_L$ are determined at $T=6$~K, and $f_H$ is determined at $T=150$~K since this peak appears only at higher temperatures.}\label{fig3}
\end{figure}

The oscillation peak discussed above was the only spectral feature observed at low temperatures. However, at higher temperatures and large currents, another peak appeared at higher frequency while the low-frequency peak became gradually suppressed [see Fig.~\ref{fig3}(a)]. The higher-frequency peak exhibited a larger linewidth and persisted to room temperature. Two spectral peaks have been previously simultaneously observed in point-contact magnetic nanooscillators~\cite{bonetti}. The low-frequency peak was identified as the self-localized spin-wave "bullet" mode, while the other peak was identified as a propagating mode within the linear spin-wave spectrum.

To establish the relation of the observed oscillation peaks to the spin-wave spectrum of the Py film, we determined the ferromagnetic resonance (FMR) frequency $f_{FMR}$ by the spin torque-driven FMR (ST-FMR) technique~\cite{fuchs}. An ac current $I_{ac}$ was applied at a microwave frequency $f_{ext}$, causing magnetization oscillation due to a combination of the Oersted field and the spin torque. The resulting periodic variation of resistance due to AMR mixed with the ac current, producing a peak of dc voltage across the sample at the value of $H$ corresponding to $f_{FMR}=f_{ext}$. Previous studies of Pt/Py bilayers showed that the ST-FMR voltage peak contains both symmetric and antisymmetric components, due to the different contributions from the spin torque and the Oersted field of the current~\cite{Liu_SHE,mosendz}. We obtained good fits to the ST-FMR peaks with a sum of symmetric and antisymmetric Lorentizan contributions [see inset in Fig.~\ref{fig3}(b)], allowing us to precisely determine $f_{FMR}$. The dependence of $f_{FMR}$ on $H$ agreed well with the Kittel formula [solid curve in Fig.~\ref{fig3}(b)]. The value $M=827$~G of Py magnetization obtained from the best fit was consistent with our magnetometry measurements and the published data for Py~\cite{Liu_SHE,mosendz}.

At $T=6$~K and $H=1.1$~kOe, the maximum frequency $f_L$ of the low-frequency peak was $0.9$~GHz below $f_{FMR}$ [Fig.~\ref{fig3}(b)]. The difference between the two frequencies decreased with decreasing $H$ to $0.5$~GHz at $H=340$~Oe.
This relationship between $f_L$ and $f_{FMR}$ was preserved at higher temperatures, confirming the findings of Ref.~\cite{Demidov_SHO} and the prediction of Ref.~\cite{slavinprl} that the auto-oscillation mode does not belong to the linear spin-wave spectrum, but instead forms a non-propagating self-localized "bullet" mode at a frequency below the spectrum.

The higher-frequency peak could be observed only at higher temperatures and large currents. Under these conditions, its frequency $f_H$ was very close to, but still remained below $f_{FMR}$, suggesting that this is also a non-propagating localized mode that does not belong to the spin-wave spectrum. Similar characteristics were exhibited by another studied device with a larger gap between the electrodes. We note that the Oersted field of the current in our devices opposes the applied field, thus reducing the frequency of the oscillation. It is possible that the high-frequency mode can become propagating if the Oersted field is reduced or reversed, for example, by utilizing a different spin Hall material such as Ta, which is characterized by the opposite sign of the spin Hall angle~\cite{Liu_Ta}. These results show that the dynamical states induced by spin current are more complicated than envisioned in the original theoretical predictions~\cite{slavinprl}, warranting further theoretical and experimental studies of self-localized dynamical modes in nanoscale magnetic systems.

\begin{figure}[htbp]
\centering
\includegraphics[width=0.485\textwidth]{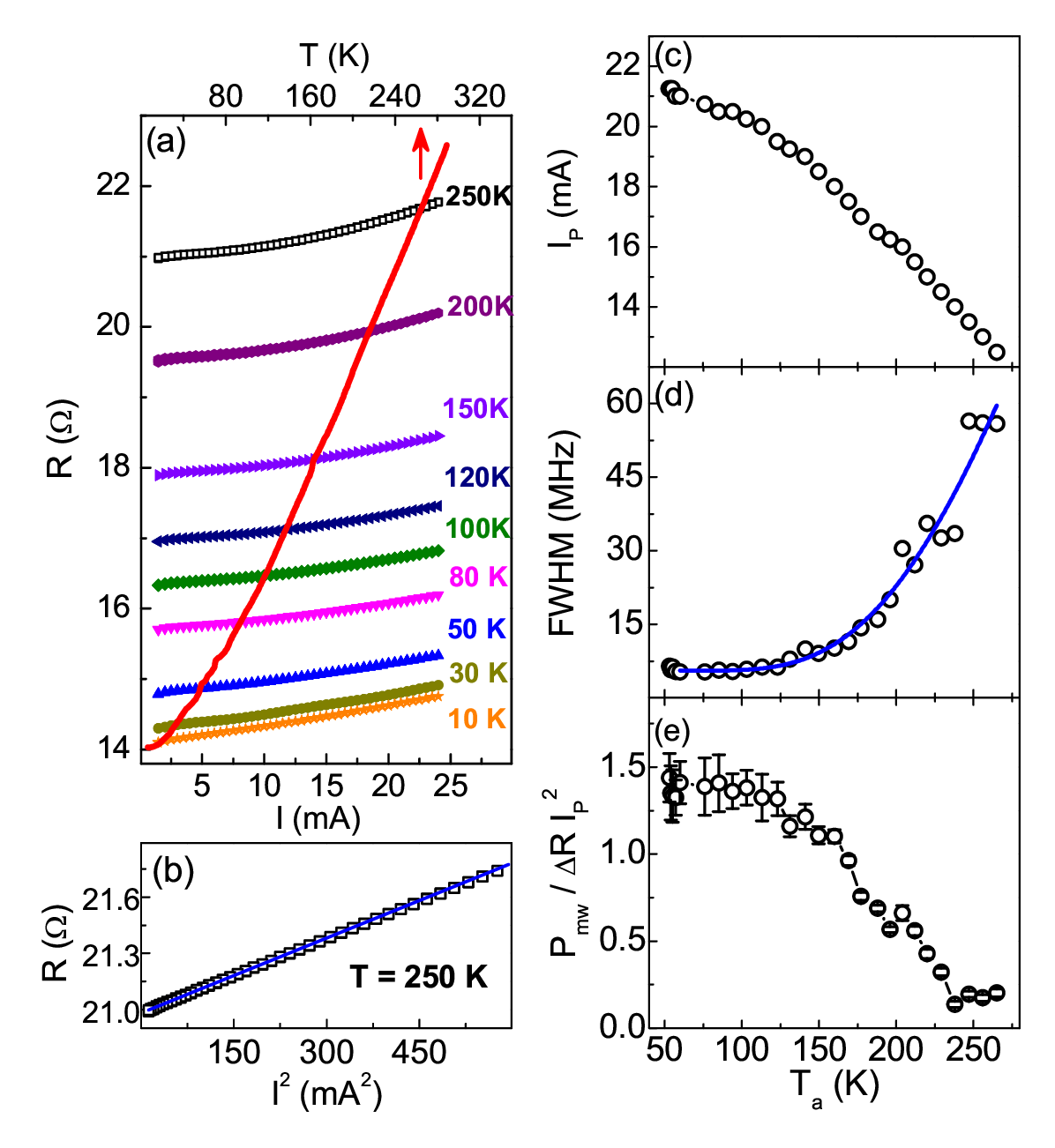}
\caption{(Color online). Effects of temperature on the spectral characteristics of the low-frequency mode. (a) $R$ {\it vs} $I$ at different substrate temperatures, as labeled, (bottom axis) and $R$ {\it vs} $T$ measured with a small ac current (top axis). (b) $R$ {\it vs} $I^2$ at 250 K. Solid line is a linear fit to the data. (c) The dependence of the current value at the peak oscillation frequency on the calculated temperature of the active device area ($T_a$). (d) Dependence of the linewidth at the peak oscillation frequency on $T_a$. Solid curves are the results of fitting by an exponential dependence $A+B*exp(-E_b/k_BT)$ with $E_b=90$~meV. (e) $P_{mw}/\Delta RI_P^2$ {\it vs} temperature, where $P_{mw}$ is the integrated microwave power under the low-frequency peak.}\label{fig4}
\end{figure}

Analysis of the temperature dependence of the oscillation provided further evidence for the bimodal nature of the dynamics induced by spin current [Fig.~\ref{fig4}]. To take into account the contribution of Joule heating to the actual temperature $T_a$ of the active device area, we first measured the dependence of the device resistance on current at different values of $T$, as illustrated by a series of curves in Fig.~\ref{fig4}(a). The dependence of resistance on current is approximately linear at low temperatures, and quadratic at high temperature[see $R$ {\it vs} $I^2$ plot in
Fig.~\ref{fig4}(b) for T = 250 K], as expected for the Joule heating effects~\cite{holm}. By comparing these dependencies to the $R$ {\it vs} $T$ dependence acquired with a small ac current that produces negligible heating [red curve and top scale in Fig.~\ref{fig4}(a)], we can obtain the dependence of $T_a$ on $I$ at a given $T$. For instance, this procedure yields $T_a=50$~K at $T=6$~K, $I=20$~mA for the oscillation spectrum shown in Fig.~\ref{fig1}(b).

We now analyze the dependence of the oscillation characteristics on the temperature $T_a$ of the active device area. The precise value of the oscillation threshold current and the oscillation characteristics at this current are difficult to determine due to the thermal noise. To avoid this difficulty, we define temperature-dependent current $I_P$ corresponding to the the highest peak generated power spectral density (PSD) [$I_P=19$~mA at $T=120$~K in Fig.~\ref{fig2}(a)]. Not only is this value well-defined based on the measured oscillation characteristics, but it is also close to the maximum of the oscillation frequency, and consequently is characterized by the smallest generation linewidth since the nonlinear line broadening effects are minimized~\cite{nonlin-oscil}, as illustrated by a dotted vertical line in Figs.~\ref{fig2}(b-d). We note that the total generated power has a maximum at a slightly larger current $I=19.5$~mA [Fig.~\ref{fig2}(d)].

Figure~\ref{fig4}(c) shows the temperature dependence of $I_P$. The value of $I_P$ decreases from $21.2$~mA at $T_a=50$~K to $12.5$~mA at $T_a=265$~K. We observed a similar temperature dependence for the threshold current. The apparent increase of the excitation efficiency at higher temperatures may be caused by a decrease of Py magnetization, an increase of the spin-current generated by SHE, and/or a decrease of the dynamical damping. Our measurements of ST-FMR did not show a significant variation of $f_{FMR}$ with temperature, indicating that the magnetization of Py does not significantly vary over the studied temperature range [see also Supplementary Materials]. We also expect that the damping of oscillation increases with increasing $T$, due to the larger contribution of nonlinear interaction of oscillation with thermal spin-waves. Therefore, the observed behaviors are likely associated with the increase of spin-current generated by SHE, warranting further studies of the temperature dependence of this effect.

We now focus our analysis on the spectral characteristics of the low-frequency peak at $I=I_P$. Figure~\ref{fig4}(d) shows the dependence of the linewidth on $T_a$ at $I=I_P$. It remains approximately constant at $T_a<150$~K, but rapidly increases at larger $T_a$. This dependence is qualitatively different from the effects of thermal fluctuations on the linewidth of a single-mode oscillation~\cite{silva,nonlin-oscil}, suggesting that the additional higher-frequency mode that appears at high temperatures and currents has a significant effect on the spectral characteristics of the oscillation. Indeed, exponential dependence of linewidth on temperature was also observed in some measurements of spin-valve nano-oscillators, and was identified with thermally activated transition between dynamical modes~\cite{sankey,schneider}. Based on the two-state fluctuation model, we were able fit the data by using $A+B*exp(-E_b/k_BT)$ with the best-fit value of $90$~meV for the energy barrier $E_b$ between the two states [curve in Fig.~\ref{fig4}(d)].

Analysis of the microwave generation power provides additional evidence for the
two-mode effects. Figure~\ref{fig4}(e) shows the temperature dependence of the integrated power $P_{mw}$ under the low-frequency peak at $I=I_P$, normalized by $\Delta RI_P^2$ to eliminate the effects of the variations of AMR and applied current on the microwave generation. This normalized quantity characterizes average oscillation amplitude of the active device area. The amplitude remains approximately constant at $T_a<120$~K, but decreases by almost an order of magnitude at $T_a=240$~K. These behaviors are correlated with the variations  of the linewidth [Fig.\ref{fig4}(e)], consistent with the decrease of the average oscillation amplitude due to the fluctuations between two dynamical modes.

To summarize, we have demonstrated emission of coherent microwaves due to the magnetization oscillation induced by a local pure spin current in a Py/Pt bilayer, which is converted into a microwave signal due to the anisotropic magnetoresistance of Py. The oscillation frequency was always lower than the frequency of the ferromagnetic resonance, confirming that the oscillation mode is a self-localized standing spin-wave. The dependence of the oscillation characteristics on current was remarkably similar to the spin-valve nano-oscillators. However, we observe a strong increase of the oscillation linewidth with temperature correlated with the emergence of an additional higher-frequency oscillation mode. These results provide insight into the  complexity of the dynamical magnetization states induced by the local spin currents, suggesting that nonlinear dynamical phenomena in nanoscale magnetic systems are still not well understood.

We thank Rongying Jin for the magnetic property measurements of the films used in our study. This work was supported by the NSF Grants ECCS-1218419 and DMR-1218414.

\end{document}